\documentclass[useAMS,usenatbib,usegraphicx,referee]{mn2e}
\usepackage{amssymb}

\def\k{km s$^{-1}$}

\def\pp{^{\prime\prime}}
\def\hi{\rm H\,{\sc i}}
\def\mjb{mJy beam$^{-1}$}
\def\jb{Jy beam$^{-1}$}

\def\deg{\hbox{$^\circ$} }

\title [G332.5--5.6, a new Galactic SNR ] {G332.5--5.6, a new Galactic
supernova remnant}
\author[Reynoso \& Green]
{E. M. Reynoso$^{1,2}$\thanks{E-mail: ereynoso@iafe.uba.ar},
\thanks{Member of the Carrera del Investigador Cient\'\i fico, CONICET, 
Argentina.} and A. J. Green$^{3}$\\
$^{1}$ Instituto de Astronom\'\i a y F\'\i sica del Espacio (IAFE),
CC 67, Suc. 28, 1428 Buenos Aires, Argentina\\
$^{2}$ Departamento de F\'\i sica, Facultad de Ciencias Exactas 
y Naturales, Universidad de Buenos Aires, Argentina\\
$^{3}$ School of Physics, University of Sydney, NSW 2006, Australia\\}

 \date{Accepted 2006.
      Received 2006;
     in original form 2006}
\begin{document}

\maketitle

\begin{abstract}
We present radio observations of the source G332.5--5.6, a candidate
supernova remnant (SNR). Observations have been performed with the 
Australia Telescope Compact Array (ATCA) at two frequencies, at
1.4 and 2.4 GHz. Our results confirm that G332.5--5.6 is an SNR, with
a spectral index $\alpha = -0.7\pm 0.2$ for the whole source and
an average fractional polarization of $\sim 35\%$ at 2.4 GHz. The 
central component is coincident with extended X-ray emission and 
the distance to the SNR is estimated to be $\sim 3.4$ kpc. Based on 
its radio and X-ray morphology, this SNR should be classified as a
composite, and we suggest that it belongs to a trident-shaped subclass 
like G291.0--0.1. 
\end{abstract}
\section{Introduction}

Supernova explosions are one of the most important sources which control the 
physical and chemical state of the interstellar medium (ISM). They inject
heavy elements into the ISM, accelerate cosmic rays, heat and compress nearby 
clouds altering the local chemistry and eventually inducing star formation.
At present, 265 supernova remnants (SNRs) are identified in our Galaxy
\citep{green06}, including those recently reported by \citet{brogan}. 
Of this sample, there are 15 with $|b|\geq 5^\circ$, but 
their distribution is assymetrical: only 4 of them are in the Southern
hemisphere, while the rest are north of declination $+5^\circ$ \citep{green}. 
Although this asymmetry is thought to be related to Gould's belt \citep[e.g.][]
{sf74}, there is no evidence of the same asymmetry in low Galactic latitude 
SNRs, which should not be statistically different \citep{green}.

\citet{dshj97} identified several SNR candidates at high Galactic 
latitude based principally on morphology at 4.85 GHz \citep{wgbe94} and the 
absence of 60-$\mu$m far-infrared emission, which is a strong discriminant 
against thermal sources. One of the most intriguing candidates found was 
G332.5--5.6, a 
region with three patches of filamentary emission and a total size 
of about $30^\prime$, that does not appear at all like a canonical SNR (a 
filamentary shell or a centrally-peaked source). The unusual appearance of 
this source meant a more comprehensive study would be needed, with spectral 
index and polarization measurements, to confirm the identification. The 
source has also been observed at 843 MHz as part of the the Sydney University 
Molonglo Sky Survey \citep[SUMSS;][]{bls99}.  

In this paper we present high resolution observations of G332.5--5.6
carried out with the Australia Telescope Compact Array (ATCA) at two
frequencies, 1384 MHz and 2432 MHz, including full Stokes parameters at the
higher frequency. 

\section{Observations and data reduction}

\begin{figure}
\includegraphics[width=252pt]{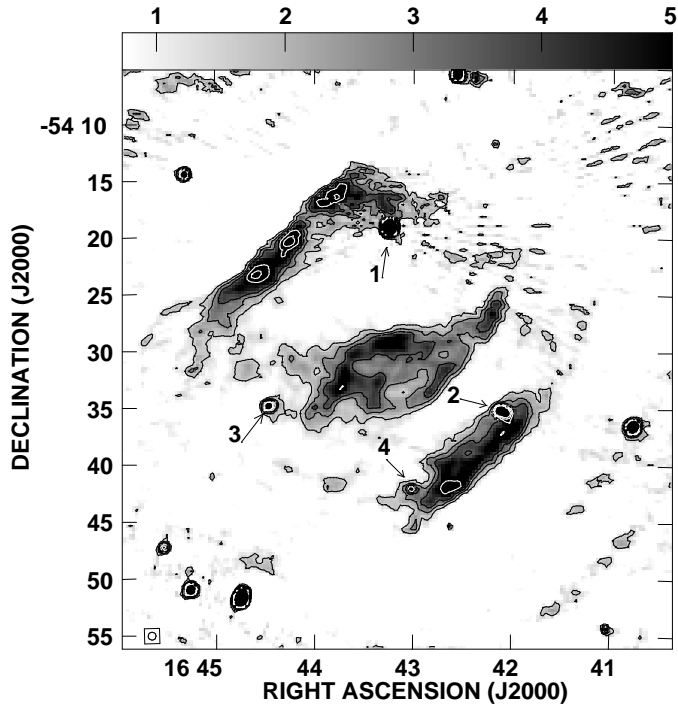}
\caption{Radio continuum mosaic of G332.5--5.6 at 1384 MHz obtained
with the ATCA. The flux density, in units of \mjb, is shown at the top of
the image. Contours have been overplotted at 1.5, 2.5, 3.5, 6, 8, 
and 10 \mjb. White lines have been used over dark regions. The
HPBW, $40\pp \times 40\pp$, is indicated in the bottom left
corner. The rms noise level is $\sim 0.5$ \mjb. The numbered arrows 
point to the compact sources listed in Table 1 and discussed 
in Section 3.2.} \label{20cm}
\end{figure}

We observed G332.5--5.6 in three sessions with the ATCA. The first 
run was on January 11 2003  over 12 hours, with the array in the 1.5B 
configuration (baselines varying from 30.6 to 1286~m), using the 
correlator configuration of two 128 MHz bands (over 32 channels) 
centered at 1384 MHz and 1704 MHz. Because of the large angular size of the
source, observations were made in mosaicing mode with three pointings 
to completely cover the extended radio emission. To optimize 
the sampling according to the Nyquist theorem, the pointings were 
separated by $19\farcm 6$. 

The remaining two sessions were conducted on  April 12 and 23 2003, 
for 5 hours each day, with the ATCA in the EW 352 configuration 
(baselines from 30.6 to 352 m). The two continuum bands were centered 
this time on 1384 MHz and 2496 MHz in the first session, and 1384 MHz 
and 2368 MHz in the second session, to avoid likely interference.
The largest angular scales sampled by these datasets are 
$35^\prime$ and $23^\prime$ at 20 cm and 13 cm respectively. The 
primary beam is smaller at 13 cm, so the observed mosaic consisted of 4 
pointings separated by $13^\prime$ from one another. Unfortunately, both 
these runs were substantially overlapped in hour angle, so the resulting
synthesized beam is rather elongated.  The source PKS 1657--56 was used to 
calibrate phase, while PKS 1934--638 was used for flux density and bandpass 
calibration. The data were processed with the MIRIAD software package
\citep{stw}. To construct the image at 20 cm, all pointings at
1384 MHz and 1704 MHz were merged together in the $uv-$plane. The
image was constructed with a cell size of $8\pp$ and uniform
weighting which minimizes the sidelobe levels. A gaussian taper 
was applied to the visibility data by setting the {\sl fwhm} parameter
equal to 24 arcsec in the MIRIAD task INVERT.

To deconvolve a dirty mosaic image, two MIRIAD routines can be
used: MOSMEM and MOSSDI. According to the MIRIAD User's Guide 
\footnote{http://www.atnf.csiro.au/computing/software/miriad/userhtml.html},
the former is generally superior, except for images
containing point sources. Therefore, the deconvolution was
performed in two steps: firstly, point sources were cleaned using
MOSSDI, and then MOSMEM was used for the residual image. The
synthesised beam was $32\pp \times 24\pp$, with a position angle 
(measured clockwise from north) of $30\deg$. However, to improve 
the signal to noise ratio, the image was smoothed to a 
$40\pp \times 40\pp$ beam.

Due to the strong artifacts introduced by the bright compact
source PMN J1643--5418, close to the western edge of the N-E
filament of G332.5--5.6, a second image was obtained repeating the
above process but excluding the 1704 MHz data. Although the
artifacts did not vanish completely, the final image was
noticeably improved. The rms noise level is $\sim 0.5$ \mjb \ 
and this image is shown in Figure \ref{20cm}. 

\begin{figure}
\includegraphics[width=252pt]{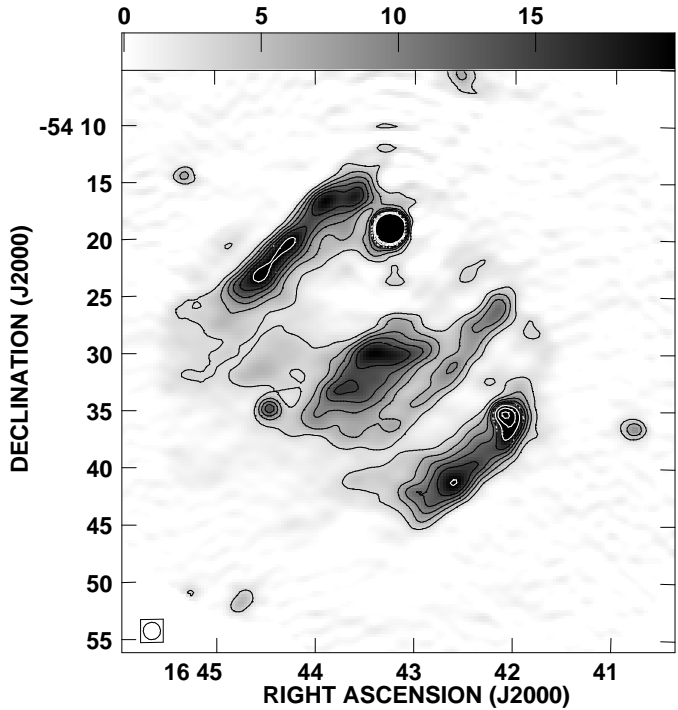}
\caption{Radio continuum mosaic of G332.5--5.6 at 2432 MHz obtained
with ATCA. The flux density, in units of \mjb, is shown at the top of
the image. Contours have been plotted at 3, 6, 9, 12, 15, 20, 25,
and 30 \mjb. White lines have been used over a dark background.
The HPBW, $90\pp \times 90\pp$, is indicated in the bottom left
corner. The rms noise level is $\sim 0.5$ \mjb.} \label{13cm}
\end{figure}

To construct the image at 13 cm, a similar procedure was
used. The observations were made with a more compact configuration 
of the ATCA, so have a lower resolution when compared with the 20 cm 
images, and a larger cell size ($15\pp$) was used. 
As before, point sources were cleaned first and the
extended emission was deconvolved using the residual image. The
clean components model was convolved with a beam of $90\pp \times
90\pp$. The rms noise of the final image, shown in Figure \ref{13cm},
is 0.5 \mjb. 

\begin{figure}
\includegraphics[width=252pt]{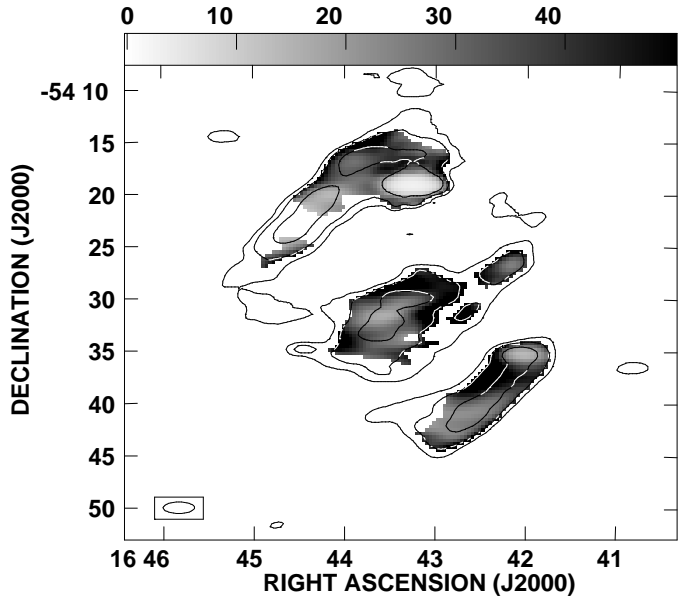}
\caption{Greyscale image of the fractional polarization of G332.5--5.6 at 
13 cm. The percentage fractional polarization level is shown on top of
the image. Continuum contours of the total intensity at 13 cm at
5, 9 and 20 \mjb \ are overlaid. The HPBW, $180\pp \times 65\pp$, 
P.A.=$-90^\circ$, is indicated in the bottom left corner.} \label{pol}
\end{figure}

The ATCA provides simultaneous recording of all four Stokes parameters 
when observing in continuum mode. We made images of the $Q$ and $U$ 
Stokes parameters at 13 cm  following the procedure described above. 
Natural weighting was used in this case and no tapering was applied
to the $uv-$data.
Both images were restored with a $180\pp \times 65\pp$ beam, and P.A.
=$-90^\circ$. A linear polarization $I_P$ image was formed by combining 
the $Q$ and $U$ images using $I_P=(Q^2+U^2)^{1/2}$. A total intensity image 
was also constructed with the same beam shape, and the fractional 
polarization was obtained by dividing $I_P$ by the total intensity image. 
Finally, the fractional polarized emission was blanked for those pixels 
where the total intensity fell below 8 \mjb \ (equivalent to $\sim 
8\sigma$) or where $I_P$ was less than 3 \mjb \ ($\sim 2\sigma$). 
In Figure \ref{pol}, the fractional polarization is shown in greyscale 
with total intensity contours overlaid. The signal-to-noise ratio
in Fig. \ref{pol} varies between 5 and 20.

\medskip

\section{Results}

Figures \ref{20cm} and \ref{13cm} display the mosaic images of
G332.5--5.6 at 20 cm and 13 cm respectively. At 20 cm, G332.5--5.6 
shows two outer ridges of emission, very straight and almost parallel, 
with a more extended central emission region elongated in the same 
direction as the two ridges. The whole structure has a trapezoidal 
shape, with the N-E and S-W outer filaments being $\sim 20^\prime$ 
and $\sim 15^\prime$ long respectively, and the distance between 
them being $\sim 30^\prime$. Both external filaments have a width 
of $\sim 4^\prime$.

The broader, central component appears to be detached from the outer 
S-W filament but could be connected with the N-E one. This central
component shows internal substructure, with a chain of knots forming 
an incomplete, elongated ring with a filament extending to the N-W. 
The filament coincides with an optical filament with characteristics 
typical of SNRs \citep{pfs04}.  There are also several compact 
sources within the area subtended by the SNR that are probably 
unrelated sources (see section 3.2). These sources are marked with 
arrows in Fig. \ref{20cm}.

The 13 cm image is very similar to the 20 cm one, although, because of
its lower resolution the central ring and Source 4 are rather blurred.

\begin{figure}
\centering
\includegraphics[width=8cm, height=9cm]{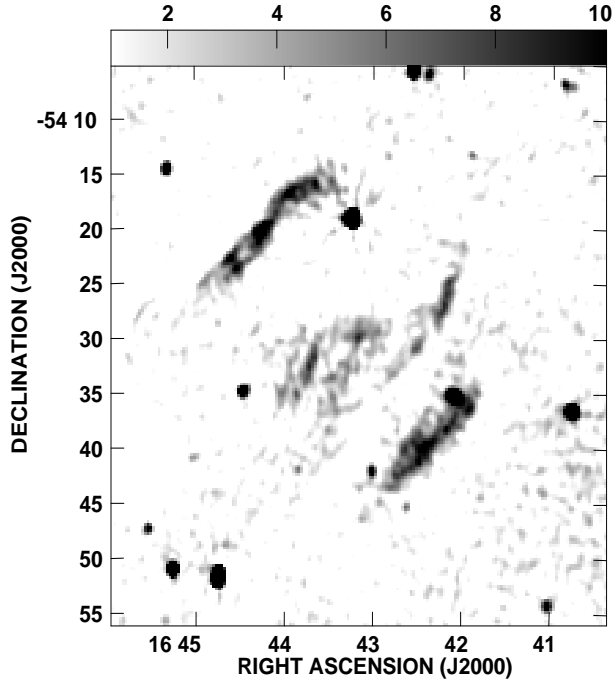}
\caption {SUMSS image of G332.5--5.6 at 843 MHz \citep{bls99}. The
flux density, in units of \mjb, is shown at the top of the image. 
The beam size is $53\pp \times 43\pp$.}\label{SUMSSimage}
\end{figure}

For comparison, in Figure \ref{SUMSSimage} we show an image of G332.5--5.6 
obtained from the Sydney University Molonglo Sky Survey 
\citep[SUMSS;][]{bls99} at 843 MHz. The morphology of the 
three filaments is very similar to the images at 13 and 20 cm.
Most of the small-scale structure observed in the 20 cm image is also 
found in the SUMSS image, including the central component,
although some of the smooth emission has been resolved out.
Moreover, at 843 MHz the two outer filaments appear relatively brighter and 
narrower than the central component.

\subsection{Polarization}

The average fractional polarization of G332.5--5.6 at 13 cm is 
$\sim 35$\%. There is little variation over the face of the SNR. 
The polarization level is rather high but \citet{dms00} also 
found high percentages of polarization in G326.3--1.8 using the
ATCA, and they discuss how the polarization might be artificially 
increased in aperture synthesis observations. This occurs
because the polarized intensity probably has more fine structure 
than the total intensity and the absence of short spacings resolves
out some of the total intensity emission.  The compact source PMN 
J1643--5418 was found to be unpolarized.

\subsection{Spectral index study}

\begin{figure*}
\includegraphics{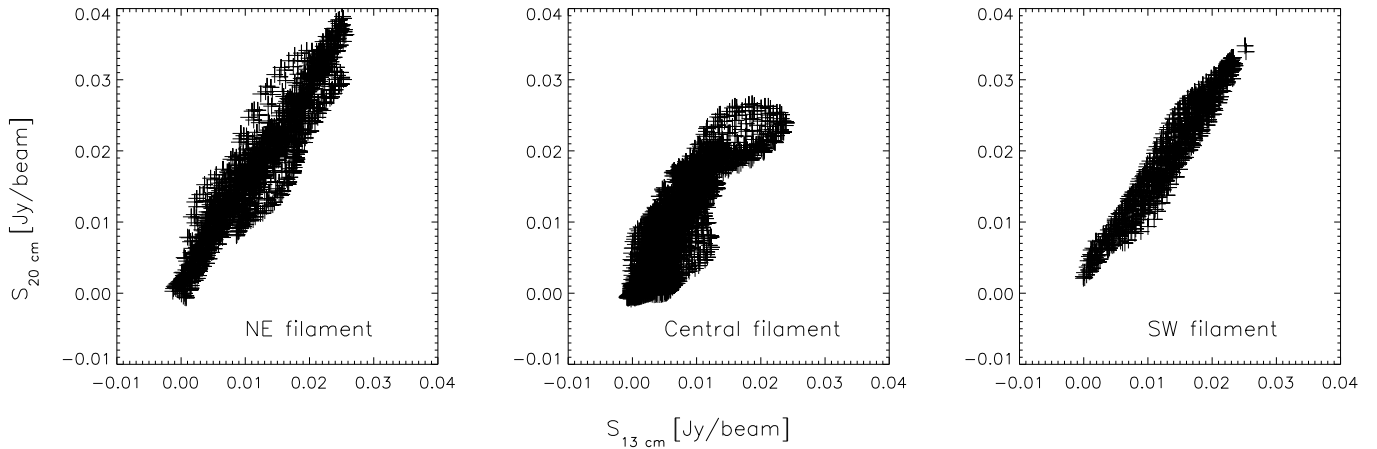}
\caption{Flux-flux plots for the three filaments of G332.5--5.6.
Flux densities at 20 cm are plotted in terms of flux densities at
13 cm. Units are in \jb.} \label{fluxfluxplot}
\end{figure*}

To obtain the spectral index of G332.5--5.6, we used a standard
flux-flux plotting technique (\citealt{costain,turtle}). This
method compares the point by point flux densities of
two images at different frequencies which have been constructed
from datasets with the same $u-v$ range. The slope of a
linear regression is then used to compute the spectral index,
independently of possible differences in the zero level of each
image.

The resolution limit of the filtered maps is determined by
the 13 cm observations. The visibilities of the two datasets
overlap between 0.3 and 1.47 k$\lambda$. Using only $u-v$ data
within this range, two images were constructed following
the same procedure as described in Section 2. The resulting
beam was $170\pp \times 63\pp$, with a position angle of
$-90^\circ$.

Flux-flux plots were constructed for the three filaments
separately. In all cases, the point sources were excluded except
for source number 4 in Table 1, which was not resolved in the
filtered maps. However, since this is a faint source, the result
is not expected to be markedly different. In fact, the S-W
filament shows the best correlation. For the other two components, 
the correlation was poor (just above 0.8). The flux-flux plots are 
shown in Figure~\ref{fluxfluxplot}. For correlations very close to 
one, the slope of the linear fit is independent of which of the 
two frequencies is taken to be the abscissa. In general, this is not 
true. Therefore it is necessary to compute two different slopes 
with each frequency being alternately the abscissa. The spectral 
indices obtained  using the average of both slopes are $-0.6 \pm 
0.3$ for the N-E filament, $-0.7 \pm 0.3$ for the central one, and $-0.5 
\pm 0.2$ for the S-W filament, where the errors enclose the values
obtained for the two different slopes. The poorest correlation is found 
for the central filament. The departures from a straight line may indicate 
that regions with different spectral indices, and hence electron populations 
with different acceleration mechanisms, are coexisting. The regions used 
in the flux density calculations were varied for a range of cutoff 
levels, to establish a measure of the uncertainty in the estimates. Higher 
resolution observations may help to undercover local differences in the 
spectral index distribution.

\begin{table}
 \centering
\begin{minipage}{140mm}
  \caption{Compact radio sources within the area of G332.5--5.6.}
  \begin{tabular}{@{}cccccl@{}}
  \hline
&&&&Flux density\\
&   R.A. (J2000)  & Decl. (J2000) & Spectral &at 20 cm\\
& h,m,s&$^\circ$ $^\prime$ $\pp$&index& (mJy)\\
\hline
1&16 43 14.6& --54 19 07&$-0.88 \pm 0.02$&$310 \pm 10$\\
2&16 42 06.4& --54 35 15&$-0.75\pm 0.15$ &$42 \pm 5$ \\
3&16 44 27.7& --54 34 53&$-0.9 \pm 0.3$&$22 \pm 2$ \\
4&16 43 01.2& --54 42 08&$-1.60 \pm 0.25$&$6.0 \pm 0.5$ \\
\hline
\end{tabular}
\end{minipage}
\begin{flushleft}
Note: Numbers correspond to Fig. \ref{20cm}. Source number 1 is 
catalogued as PMN 1643--5418.\\
\end{flushleft}
\label{tab1}
\end{table}

To study the compact sources with a better resolution than that
provided by the filtered images at 13 and 20 cm, we applied the
flux-flux method combining the MOST survey image at 843 MHz and
our original 20 cm image, convolved to the resolution of the
Molonglo telescope ($53\pp \times 43\pp$). This comparison
is only valid for compact sources, since the Molonglo telesope
filters out smooth extended structures that may be detected with
the ATCA. In Table 1 we show the spectral indices computed using 
the higher resolution data. From the steep negative indices measured, 
it is likely that all four sources are extragalactic. 

We have also measured the integrated flux density of the three filaments 
at 13 and 20 cm. The results are listed in columns 2 and 3 of Table 2. 
To estimate the integrated flux density from the remnant, including those 
regions where there are unrelated sources superimposed (e.g. in the 
S-W filament), the integrated flux densities of these sources were 
computed and subtracted.  Spectral indices computed by direct application 
of the relation $S_1/S_2 = (\nu_1/\nu_2)^\alpha$ using these integrated 
flux densities  are listed in column 4. These indices are in good agreement 
with those obtained with the flux-flux method (last column in Table 2). The 
total flux density of G332.5--5.6 is estimated to be $1.90\pm 0.15$ 
Jy at 20 cm, and $1.3\pm 0.2$ Jy at 13 cm. Combining these total flux
densities, the global spectral index for the source is estimated to be 
$-0.7\pm 0.2$.

\medskip
\subsection{The distance to G332.5--5.6}

To analyze the nature of the SNR, it is crucial to
know its distance. A very powerful tool to constrain distances to 
radio sources is to analyze the \hi~$\lambda$21 cm line in absorption 
against the continuum emission. To construct an 
absorption profile towards G332.5--5.6, we made use of \hi~ observations 
from the Southern Galactic Plane Survey \citep[SGPS;][]{sgps}. The 
SGPS combines observations from the ATCA and the 64 m Parkes telescope. 
At the high Galactic latitude of this object, where only Parkes data
are available, the observations have a resolution of $16'$ and a
sensitivity of $\sim 300$ mK.

\begin{figure}
\includegraphics[width=252pt]{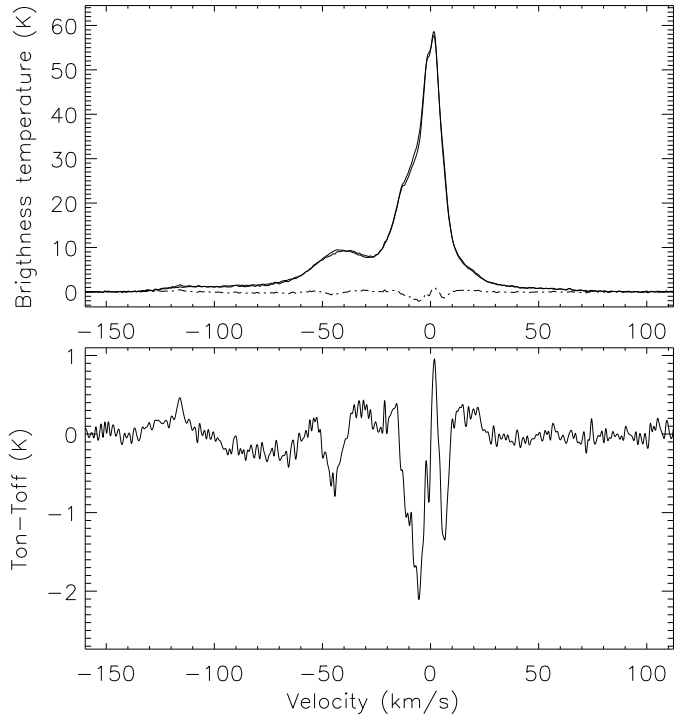}
\caption{Top: \hi~ profiles on- and off-source (solid lines) and the
difference $T_{on} - T_{off}$ (dashed line). Bottom: Zoomed-in 
difference profile. All velocities are referred to the LSR.}
\label{HIprofile}
\end{figure}

The radio continuum image at 20 cm was convolved to the same beam
as the \hi~data. This image was used as a template to compute two
\hi~data cubes: one with all pixels blanked except where the continuum 
emission was above 0.5 \jb, and the other blanking only the pixels 
inside the same region. The first cube was used to obtain the on-source 
profile, while the off-source profile was produced from 
the average \hi~ emission in the surroundings of the blanked
region in the second cube. The absorption profile, shown in
Figure~\ref{HIprofile}, is the difference between the two profiles. 
This assumes that the emission profile is not varying significantly over 
the region of the SNR.

\begin{table}
 \centering
\begin{minipage}{140mm}
  \caption{Measured radio properties for G332.5--5.6 components.}
  \begin{tabular}{@{}lcccc@{}}
  \hline
& Flux density&Flux density\\
&at 1384 MHz&at 2432 MHz&Spectral&Spectral\\
Filament&(Jy)&(Jy)&index $\alpha^a$&index $\alpha^b$\\
\hline
N-E&$0.63 \pm 0.05$&$0.43 \pm 0.03$&$-0.7\pm 0.3$&$-0.6\pm 0.3$\\
Central&$0.73\pm 0.05$&$0.50\pm 0.04$&$-0.7\pm 0.3$&$-0.7\pm 0.3$\\
S-W&$0.43\pm 0.03$&$0.30\pm 0.02$&$-0.6\pm 0.3$&$-0.5\pm 0.2$\\
\hline
\end{tabular}
\end{minipage}
\begin{flushleft}
{$^a$ {${S_1/ S_2} =({\nu_1 / \nu_2})^\alpha$} \\
$^b$ Flux-flux method.}\\
\label{tabfil}
\end{flushleft}
\end{table}

The optimal procedure with single dish data is to obtain the on-source 
profile from a pointing towards the peak of the continuum source, while 
the off-source profile is usually obtained from an average of pointings
surrounding the source. Since we are using a published archive, pointings 
are not expected to be optimal. However, since G332.5--5.6 is about twice
as large as the Parkes beam, it is likely that the profiles used
are substantially independent.  In spite of some drifts in the zero level 
due to uncertainties in the method, three absorption features can be 
reliably identified. Two of them, 
near 0 \k, are related to local gas, and the broader one includes a 
secondary minimum at about $-10$ \k \ corresponding to the Sagittarius 
arm \citep[][with R$_{\odot}=8.5$ kpc]{gg76}. The emission excess at
0 \k \ can be explained as background \hi~ from the far side of
the Scutum-Crux arm. The third absorption feature is centred at
approximately $-45$ \k, with a FWHM of ~7 \k, and represents the near
side of the Scutum-Crux arm. Adopting a lower limit of $-50$
\k \ for the systemic velocity of the SNR (allowing for the
width of the --45 \k \ feature), the Galactic rotation model of
\citet{fbs} produces distances of 3.4 and 12 kpc. The emission
excess at the tangent point, near $-115$ \k, resolves the ambiguity
in favor of  the closer distance. The remnant would also be extraordinarily
large ($\sim 100$ pc) and unreasonably high above the Plane ($> 1$
kpc) at the far distance.


To confirm our absorption result, it would be useful to observe
this source with the ATCA in the \hi \ 21 cm line, since interferometers 
filter extended emission and thus the absorption profile is 
directly observed. In addition, the high angular resolution that 
can be achieved with the ATCA will distinguish between absorption 
from G332.5--5.6 and that produced by the brighter compact source PMN
J1643--5418. At present, such a detailed study is not possible.

\medskip
\section{Discussion}

\subsection{Comparison with X-rays data}

\begin{figure}
\centering
\includegraphics[width=8.1 cm, height=8.1 cm]{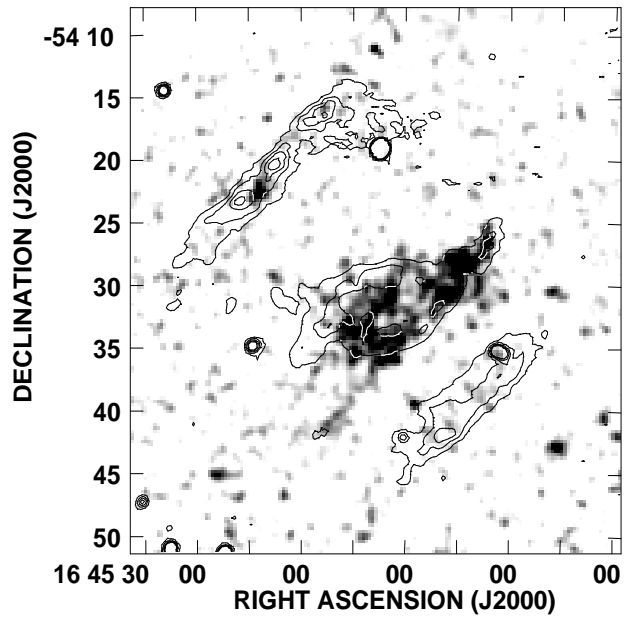}
\caption {ROSAT PSPC image in greyscale, smoothed with a $40\pp$ 
gaussian. Contours of the SNR at 20 cm are overlaid.} \label{rosat}
\end{figure}

We have searched for an X-ray counterpart of G332.5--5.6 in the ROSAT 
PSPC catalogue. These data have a nominal resolution of $\sim 30\pp$.
To improve the sensitivity, we have smoothed the image with a $40\pp$
Gaussian. In Figure \ref{rosat} we show the X-ray emission in 
the direction of the SNR with radio contours at 20 cm overlaid. 

There is strong X-ray emission coming from the center of the 
remnant, while no significant X-ray emission is found at the
outer filaments except for a small source in the northern filament, 
in the center of two radio peaks. This source is catalogued as 
J164422.2--542221 in the Second ROSAT PSPC Catalog. However, the 
association between this X-ray source and the SNR is uncertain.

The central component and the X-ray emission
are morphologically very similar. There is a weak finger in X-rays
extending approximately $7^\prime$ to the SE beyond the radio
outermost contour. 
 
The PSPC image covers the frequency range from 0.1 to 2.4 keV.
From the ROSAT All Sky Survey (RASS3) images, it is clear that most
of the X-ray emission arises in the frequency range from 0.5 to
2.4 keV, while the emission between 0.1 and 0.5 keV is negligible.

The central component does not have a flat spectral index, compared
with the two outer filaments, which would be characteristic of
recent energy injection. Hence, our result does not support a model where
the central X-ray emission is associated with a nebula driven by the 
winds of a central pulsar, a ``pulsar wind nebula'' (PWN), which is 
characterized by flat spectra \citep[e.g.][]{elsapwn}
 
However, as noted in section 3.2, the poor correlation between the emission 
at 13 and 20 cm for the central component may imply that different electron 
populations on smaller scales are coexisting, and thus a PWN cannot be 
definitely ruled out. To investigate this possibility, we searched the ATNF 
Pulsar Catalogue 
\citep{man05}\footnote{http://www.atnf.csiro.au/research/pulsar/psrcat}. 
No pulsars were found in the vicinity of G332.5--5.6. Hence, we believe 
that the central X-ray emission is produced by an alternative mechanism, 
such as cloud evaporation in the hot SNR interior \citep{wl91} or by thermal
conduction, when the shell becomes radiative and only the hot interior persists
in X-ray \citep[e.g.][]{har97}. An improved  ROSAT X-rays image which 
may help to discriminate between the different scenarios is presented by 
\citet{milorad}. Additionally, since the Parkes Multibeam Pulsar Survey
only extends to $|b| < 5^\circ$, a search for a pulsar towards G332.5--5.6
should be pursued.

\medskip

\subsection{The nature of G332.5--5.6}

The morphology of G332.5--5.6 and its linear size of $\sim 30$ pc, 
estimated for a distance of 3.4 kpc, indicates a more evolved SNR.
The source resembles bilateral SNRs, like G350.0--02.0 and VRO 42.05.01. 
This class of SNRs is characterised by a clear axis of symmetry with two 
bright limbs and occasionally low levels of emission along the axis 
\citep{gans98}. Several explanations have been proposed to account for 
such morphology \citep[e.g.][]{bks95}. The most plausible appears to be 
that these remnants are expanding into elongated cavities in the interstellar 
medium, which may have been created by the ambient magnetic field 
\citep{gans98}.

In general, the symmetry axes of bilateral SNRs are oriented nearly
parallel to the Galactic plane. The axis of G332.5--5.6, on the
contrary, is almost perpendicular to this direction. The same
``anomalous'' orientation is seen in the two bilateral SNRs: G296.5+10.0
and G327.6+14.6 (SN 1006). These two remnants are also exceptional in 
their height above the Galactic plane. Using the distance estimated by
\citet{gdggg00} based on \hi \ observations, G296.5+10.0 lies
$350^{+300}_{-150}$ pc above the Galactic disk, while SN 1006 has
a height of $570\pm 20$ pc based on the distance derived by \citet{wgl03}
combining proper motions of optical filaments and radial velocities.

\begin{figure}
\centering
\includegraphics[width=8.1 cm, height=8.1 cm]{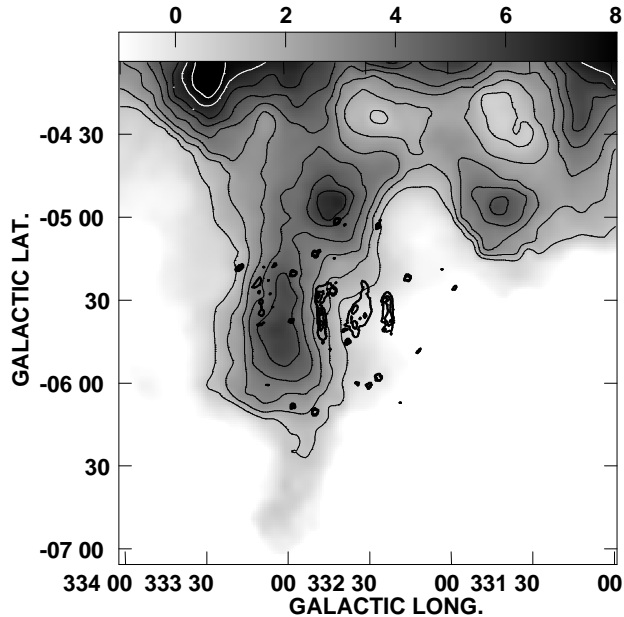}
\caption {\hi \ brightness temperature distribution, in greyscale and 
contours, averaged between --41.8 and --51.7 \k \ towards a broad region 
around G332.5--5.6 \citep[SGPS;][]{sgps}. The greyscale in K, is indicated 
at the top of the image. The \hi \ contours range from 1 to 8 K, in steps 
of 1 K. A few thick, black contours of the radio continuum 
emission at 20 cm are shown.} \label{HImap}
\end{figure}

At a distance of 3.4 kpc, G332.5--5.6 would be located at a height of 330 pc 
above the Plane. \citet{gans98} explains the orientation in these cases as 
due to perpendicular structures such as chimneys and ``worms'', which occur
frequently at high Galactic latitudes \citep[e.g.][]{heil84}. The \hi \
distribution between about $-45$ and --50 \k, which encompasses the 
systemic velocity of the remnant, shows a protrusion of enhanced density 
that extends from the Galactic plane to the halo. The N-E filament of the SNR 
appears roughly aligned, near this protrusion. Figure \ref{HImap} shows 
the average \hi \ emission within 10 \k \ around $v=-46$ \k, where the 
density enhancement is well defined. It is not clear whether the match 
between the N-E filament and the \hi \ emission is due to absorption effects 
or to a physical interaction between the SNR shock front and the ISM. There 
is no evidence of an association between the S-W filament and the \hi \ gas.

In fact, rather than a bilateral SNR, G332.5--5.6 resembles a trident
like G291.0--0.1 \citep{wg96}. Both SNRs are composed of three roughly 
parallel ridges, perpendicular to the Galactic disk, and both have a bright 
X-ray core coincident with the central radio component. For 
G291.0--0.1 this central region is a PWN \citep{har98} and there is also 
a circular region of fainter emission. However, for G332.5--5.6, which has a 
larger angular size, there is no indication of extended underlying emission 
although a smooth component could have been filtered out because 
interferometry was used for the observations. Single dish measurements will 
be needed to address this question. Another difference between these two 
SNRs is that for G291.0--0.1, the central component is more intense and 
extended than the outer filaments, while for G332.5--5.6 all three components 
are roughly equal in size and intensity. 

Although the outer filaments may be part of a shell structure, 
\citet{r+86} propose that all the components of G291.0--0.1 are  
subsidiary peaks of a filled-centre SNR. Nevertheless, we believe despite 
the differences described, the morphological radio and
X-rays coincidences between these remnants are so striking that they
may well be members of a new, trident-shaped, subclass of composite SNRs.

\section{Conclusions}

We have observed the radio source G332.5--5.6 with the ATCA at two radio
continuum frequencies, 1384 MHz and 2432 MHz. Our observations confirm
that this source is a SNR. On average, the emission from the source is about 
35\% polarized. Using these two frequencies, the global spectral index for 
the SNR was found to be $\alpha = -0.7\pm 0.2$. All of the three main 
components of the SNR have essentially the same spectral characteristics.
However, the departures from a linear correlation
in the flux-flux plots for the central filament indicate that probably
more than one population of electrons is present. Therefore, the
existence of a pulsar wind driven nebula cannot be ruled out at this
stage. The central component is coincident with extended X-ray emission. 
A more detailed spectral study of this X-ray component would be helpful
in determining the origin of this emission. 

We have used \hi \ archival data to analyze the absorption spectrum
towards this source. Based on this spectrum, we found that the distance 
to the SNR is $\sim 3.4$ kpc. At this distance, the height above the
Galactic disk is 330 pc. 

From its peculiar radio and X-ray morphology, G332.5-5.6 should be 
classified as a composite SNR, and we suggest that it belongs to a 
trident-shaped subclass with G291.0--0.1. The orientation of the three 
components is perpendicular to the Galactic plane, similar to that found
for the bilateral SNRs G296.5+10.0 and SN 1006, which also lie well above 
the Galactic plane (heights $> 300$ pc). This orientation could be related 
to chimneys and ``worms'' in the halo. We found an \hi \ feature extending 
towards the halo at a distance of 3.4 kpc (systemic velocity of  
$\sim -50$ \k) which appears aligned with the SNR, but the evidence for an 
association is weak.

\section*{Acknowledgements}

This project was partially financed by grants ANPCyT-14018, PIP-CONICET 6433 
and UBACYT A055 (Argentina). During part of this work, EMR was a visiting 
scholar at the University of Sydney. The ATCA is part of the Australia 
Telescope, which is funded by the Commonwealth of Australia for operation as
a National Facility managed by CSIRO. The Molonglo Observatory is owned and
operated by the University of Sydney. We have made use of the ROSAT Data 
Archive of the Max-Planck-Institut f\"ur extraterrestrische Physik (MPE) 
at Garching, Germany.


\begin{thebibliography}{}

\bibitem[\protect\citeauthoryear{Bisnovatyi-Kogan \& Silich}{1995}]{bks95}
Bisnovatyi-Kogan, G. S., Silich, S. A., 1995, RvMP, 67, 661

\bibitem[\protect\citeauthoryear{Bock, Large \& Sadler}{1999}]{bls99}Bock,
D. C.-J., Large, M. I., Sadler, E. M., 1999, AJ, 117, 1578

\bibitem[\protect\citeauthoryear{Brogan et al.}{2006}]{brogan}Brogan, C. 
L., Gelfand, J. D., Gaensler, B. M., Kassim, N. E., Lazio, T. J. W., 2006,
ApJL, 639, 25

\bibitem[\protect\citeauthoryear{Costain}{1960}]{costain}
Costain C. H., 1960, MNRAS, 120, 248

\bibitem[\protect\citeauthoryear{Dickel, Milne \& Strom }{2000}]{dms00}
Dickel, J. R., Milne, D. K., Strom, R. G., 2000, ApJ, 543, 840

\bibitem[\protect\citeauthoryear{Duncan et al.}{1997}]{dshj97}
Duncan A.R., Stewart R.T., Haynes R.F., Jones K.L., 1997, MNRAS,
287, 722

\bibitem[\protect\citeauthoryear{Fich, Blitz \& Stark}{1989}]{fbs}
Fich M., Blitz L., Stark A. A., 1989, ApJ, 342, 272

\bibitem[\protect\citeauthoryear{Gaensler}{1998}]{gans98}
Gaensler, B. M., 1998, ApJ, 493, 781

\bibitem[\protect\citeauthoryear{Georgelin \& Georgelin}{1976}]{gg76}
Georgelin Y. M., Georgelin Y. P., 1976, A\&A, 49, 57

\bibitem[\protect\citeauthoryear{Giacani et al.}{2000}]{gdggg00}Giacani,
E. B., Dubner, G. M., Green, A. J., Goss, W. M., Gaensler, B. M., 2000,
AJ, 119, 281

\bibitem[\protect\citeauthoryear{Giacani et al.}{2001}]{elsapwn}Giacani,
E. B., Frail, D. A., Goss, W. M., Vieytes, M., 2001, AJ, 121, 3133

\bibitem[\protect\citeauthoryear{Green}{2004}]{green}
Green, D. A., 2004, Bull. Astr. Soc. India, 32, 335

\bibitem[\protect\citeauthoryear{Green}{2006}]{green06}Green, D. A., 
 2006, ``A Catalogue of Galactic Supernova Remnants (2006 April version)'', 
Astrophysic Group, Cavendish Laboratory, Cambridge, United Kingdom available 
at ``http://www.mrao.cam.ac.uk/surveys/snrs/'' 

\bibitem[\protect\citeauthoryear{Harrus et al.}{1997}]{har97}
Harrus, I. M., Hughes, J. P., Singh, K. P., Koyama, K., Asaoka, I., 
1997, ApJ, 488, 781

\bibitem[\protect\citeauthoryear{Harrus et al.}{1998}]{har98}
Harrus, I. M., Hughes, J. P., Slane, P. O., 1998, ApJ, 499, 273

\bibitem[\protect\citeauthoryear{Heiles}{1984}]{heil84}
Heiles, C., 1984, ApJS, 55, 585

\bibitem[\protect\citeauthoryear{Manchester et al.}{2005}]{man05}
Manchester R. N., Hobbs G. B., Teoh A.,  Hobbs M., 2005, AJ, 129,
1993

\bibitem[\protect\citeauthoryear{McClure--Griffiths et al.}{2005}]{sgps}
McClure--Griffiths N. M., Dickey J. M., Gaensler B. M., Green A.
J., Haverkorn M., Strasser S., 2005, ApJS, 158, 178

\bibitem[\protect\citeauthoryear{Parker, Frew \& Stupar}{2004}]{pfs04}
Parker Q.A., Frew D.J., Stupar M., 2004, AAO Newsletter, 104, 9

\bibitem[\protect\citeauthoryear{Roger et al.}{1986}]{r+86}
Roger, R. S., Milne, D. K., Caswell, J. L., Little, A. G., 1986,
MNRAS, 219, 815

\bibitem[\protect\citeauthoryear{Sault, Teuben \& Wright}{1995}]{stw}
Sault R.  J., Teuben P. J., Wright M. C. H., 1995, in ASP Conf.
Ser. 77, Astronomical Data Analysis Software and Systems IV, ed.
R. A. Shaw, H. E. Payne, \& J. J. E. Hayes (San Francisco: ASP),
433

\bibitem[\protect\citeauthoryear{Stothers \& Frogel}{1974}]{sf74}
Stothers, R., Frogel, J. A., 1974, AJ, 79, 456

\bibitem[\protect\citeauthoryear{Stupar et al.}{2006}]{milorad}
Stupar, M., et al., 2006, in preparation

\bibitem[\protect\citeauthoryear{Turtle et al.}{1962}]{turtle}
Turtle A. J., Pugh J. F., Kenderdine S., Pauliny-Toth I. I. K.,
1962, MNRAS, 124, 297

\bibitem[\protect\citeauthoryear{White \& Long}{1991}]{wl91}
White, R. L., Long, K. S., 1991, ApJ, 373, 543

\bibitem[\protect\citeauthoryear{Whiteoak \& Green}{1996}]{wg96}Whiteoak,
J. B. Z., Green, A. J., A\&ASS, 118, 329

\bibitem[\protect\citeauthoryear{Winkler, Gupta \& Long}{2003}]{wgl03}
Winkler, P. F., Gupta, G., Long, K. S., 2003, ApJ, 585, 324

\bibitem[\protect\citeauthoryear{Wright et al.}{1994}]{wgbe94}
Wright, A. E., Griffith, M. R., Burke, B. F., Ekers, R. D.,
1994, ApJS, 91, 111

\end{thebibliography}
\end{document}